\documentclass{JHEP3}
\usepackage{units}
\usepackage{amsmath}
\usepackage{epsfig}
\newcommand{\be}{\begin{equation}}
\newcommand{\ee}{\end{equation}}
\newcommand{\ba}{\begin{eqnarray}}
\newcommand{\ea}{\end{eqnarray}}

\title{Quantum scale-invariant models as effective field theories}

\author{M. E. Shaposhnikov\\
  Institut de Th\'eorie des Ph\'enom\`enes Physiques,\\
  \'Ecole Polytechnique F\'ed\'erale de Lausanne,\\
  CH-1015 Lausanne, Switzerland\\
  E-mail: \email{Mikhail.Shaposhnikov@epfl.ch}
  }

\author{F.V. Tkachov\\
  Institute for Nuclear Research of the Russian Academy of Sciences,\\
  60th October Anniversary prospect 7a, Moscow 117312, Russia\\
  E-mail: \email{ftkachov@ms2.inr.ac.ru}}

\abstract{We address the question of whether the quantum
scale-invariant theories introduced in \cite{Shaposhnikov:2008xi} are
renormalizable or play the role of effective field theories that are
valid below the Planck scale $M_P$. We show that starting from
two-loop level the renormalization procedure requires introduction of
counter-terms with structures different from those in the initial
Lagrangian, making these theories non-renormalizable and therefore
non-predictive above $M_P$. Despite non-renormalizability, the
attractive features of these theories, associated with the stability
of the Higgs mass agains radiative corrections and the smallness of
the cosmological constant, remain intact.} 

\date{\today}

\begin{document}
%%%%%%%%%%%%%%%%%%%%%%%%%%%%%%%%%%%%%%%%%%%%%%%%%%%%%%%%%%%%%

Ref. \cite{Shaposhnikov:2008xi} introduced a class of models with
scale-invariance at the quantum level. Quantum scale invariance (QSI)
would forbid all mass parameters in the Lagrangian, including the
Higgs mass and cosmological constant, while the spontaneous breaking
of QSI introduces all mass scales, including the gravitational
constant. Some intriguing phenomenological consequences of these
theories were discussed in  \cite{Shaposhnikov:2008xb}. They are
related to the stability of the Higgs mass agains radiative
corrections, to the cosmological inflation, and existence of dark
energy. The key point  of phenomenological considerations of
\cite{Shaposhnikov:2008xb,Shaposhnikov:2008xi} is related to the
existence of exact symmetry and its spontaneous breakdown. Though the
consequences of spontaneously broken QSI theories remain valid
independently of the character of these models (renormalizable
theories versus effective field theories, valid up to the Planck
scale  $M_P\sim 10^{19}$ GeV), it is interesting to clarify whether
these models are renormalizable or can only play the role of effective
field theories.

A simple toy scalar field model considered in
\cite{Shaposhnikov:2008xi} is given  by the action
\begin{eqnarray}
\nonumber
&&{S}= \int d^4x \; {\cal L},\\ 
 \label{L}
&&{\cal L}=\frac{1}{2}\left[(\partial_\mu\chi)^2 +(\partial_\mu h)^2
\right]-V(h,\chi),\\
&&V(h,\chi)= \lambda\left(h^2-\zeta^2\chi^2\right)^2.
 \nonumber
\end{eqnarray}
It is scale-invariant at the classical level. The potential is chosen
in such a way that the scale invariance is spontaneously broken along
the flat directions $h=\pm\zeta\chi$.  The fields  $h$ and $\chi$ can
be thought of as the Higgs field and the dilaton, correspondingly.
Phenomenological considerations require that one should take  $\zeta
\sim v/M_{\rm P}\sim 10^{-17} \lll 1$, where $v\sim 100$ GeV is the
electroweak scale.

To get a scale invariant theory on the quantum level one can replace
the  parameter $\mu$ of dimensional regularization (we take the
dimensionality of space-time equal to $n=4-2\epsilon$) by an
appropriate combination of dynamical fields $\chi$ and $h$, and remove
divergences by droping the contributions that are singular in the
limit $\epsilon\to 0$ \cite{Shaposhnikov:2008xi}. This procedure was
suggested already in  \cite{Englert:1976ep}, see also
\cite{Jain:2008qv}. A choice, motivated by  cosmological
considerations \cite{Shaposhnikov:2008xb}, reads
\be
\mu^{2\epsilon} \to \left[\omega^2\right]^{\frac{\epsilon}{1-\epsilon}}~,
\label{GR}
\ee
where $\omega^2 \equiv \left(\xi_\chi \chi^2 + \xi_h h^2\right)$ and 
$\xi_\chi,~\xi_h$ are the couplings of the scalar fields to the
gravitational Ricci scalar. The low energy theory contains a massive
Higgs field and the massless dilaton.

In this short note we argue that these types of models  are
non-renormalizable and thus require an ultraviolet (UV) completion, at
least if the standard way of renormalization is used.

To demonstrate that this is indeed the case it is sufficient to
consider the simplest massless scalar model with a quartic
self-interaction. It can be seen that the non-renormalizability
argument holds in the more general case. We use the notations of Sec.
2 of \cite{Shaposhnikov:2008xi} and, for convenience, start from  the
dimensionally regularized bare action in the form 
\be
\label{simple}
S = \int {d^n x} \left[ {\left( {\partial _\mu  \varphi _B } 
\right)^2  - \mu ^{2\varepsilon } \lambda _B \varphi _B ^4 } \right]~.
\ee
First reprepesent the action in terms of finite fields: 
\be
\varphi _B \left( x \right) = \mu ^{ - \varepsilon} 
Z_\varphi  ^{\frac{1}{2}} \varphi (x)~,
\ee
where the standard field renormalization factor is introduced. 
The action becomes 
\be
S = \int {d^n x\mu ^{ - 2\varepsilon} } 
\left[ {Z_\varphi  \left( {\partial _\mu  \varphi (x)} \right)^2  
- \lambda _B Z_\varphi  ^2 \varphi ^4 (x)} \right]~.
\ee
With a proper choice of the  $Z$'s and  $\lambda _B$, the theory
becomes finite order-by-order in powers of the coupling  $\lambda$.
The simplest variant of the construction of Sec. 2 of
\cite{Shaposhnikov:2008xi} is to replace 
\be
\mu ^2  = \mu _0 ^2  + \xi _B \varphi ^2 \left( x \right)~.
\ee 
Within perturbation theory, perform the expansion around  $\mu _0 ^2
$:
\be
S = \int {d^D x\mu _0 ^{ - 2\varepsilon } }  
\left[ {Z_\varphi  \left( {\partial _\mu  \varphi (x)} \right)^2  - 
\lambda _B Z_\varphi  ^2 \varphi ^4 (x) - 
\varepsilon \mu _0 ^{ - 2} 
\xi _B Z_\varphi  
\left( {\partial _\mu  \varphi (x)} \right)^2 
\varphi ^2 \left( x \right) + O\left( {\varphi ^6 } \right)} \right]~.
\ee
One notices the presence of an abnomal vertex (the one with two
derivatives and the factor  $\varepsilon $ in the coupling) along with
the normal one  (without derivatives, and with the usual coupling 
$\lambda $).

Since each UV-divergent loop in a dimensionally regulated perturbative
integral (diagram) gives one factor  $\varepsilon^{-1}$, whereas each
abnormal vertex adds the factor  $\varepsilon$, potentially
interesting integrals must have the number of loops larger by one or
more than the number of abnormal vertices. In a normal scalar model
(no abnormal vertices), the 4-leg one-particle-irreducible diagrams
diverge logarithmically. Each abnormal vertex adds a second power of
momenta to the numerator of the integrand of the diagram, so that the
index of UV divergence is increased by two. Since the action already
contains a vertex with two derivatives, such divergences can be
absorbed into the factor $\xi_B$. Therefore, problematic divergences
can emerge only in diagrams with two or more abnormal vertices. 

We conclude that although there are no problems with UV
renormalization of the  model (\ref{simple}) at the one- and two-loop
level, there are non-renormalizable UV divergences starting from three
loop level upwards. One can see that the same conclusion holds for
more complex models (non-zero masses, more fields, shifts of the
fields by c-number constants). A similar argument in the case of
fermions Yukawa-coupled to scalars, establishes that the
nonrenormalizable contributions begin to occur at the two-loop level;
the same is true for scalar theories with massive fields. If one
introduces gauge fields into the model, then the one-loop
renormalizability argument of \cite{Shaposhnikov:2008xi} remains
correct, whereas starting from two-loop level, non-renormalizable UV
divergences start to occur via the outlined mechanism.

Let us turn now back to the Lagrangian (\ref{L}). In the Higgs sector
(no dilatons in external legs, which are suppressed by the small value
of  $\zeta$) the contribution which is leading in momenta  comes from
the three-loop graph shown in Fig. \ref{3loop}. It contains two normal
vertices $\lambda h^4$ and one abnormal vertex, coming from the
expansion of $\mu$ with respect to $h$ in (\ref{GR}). 

%%%%%%%%%%%%%%%%%%%%%%%%%%%%%%%%%%%%%%%%%%%%%%%%%%%%%%%
\FIGURE{
\hspace*{2.3cm}\includegraphics[width=7cm]{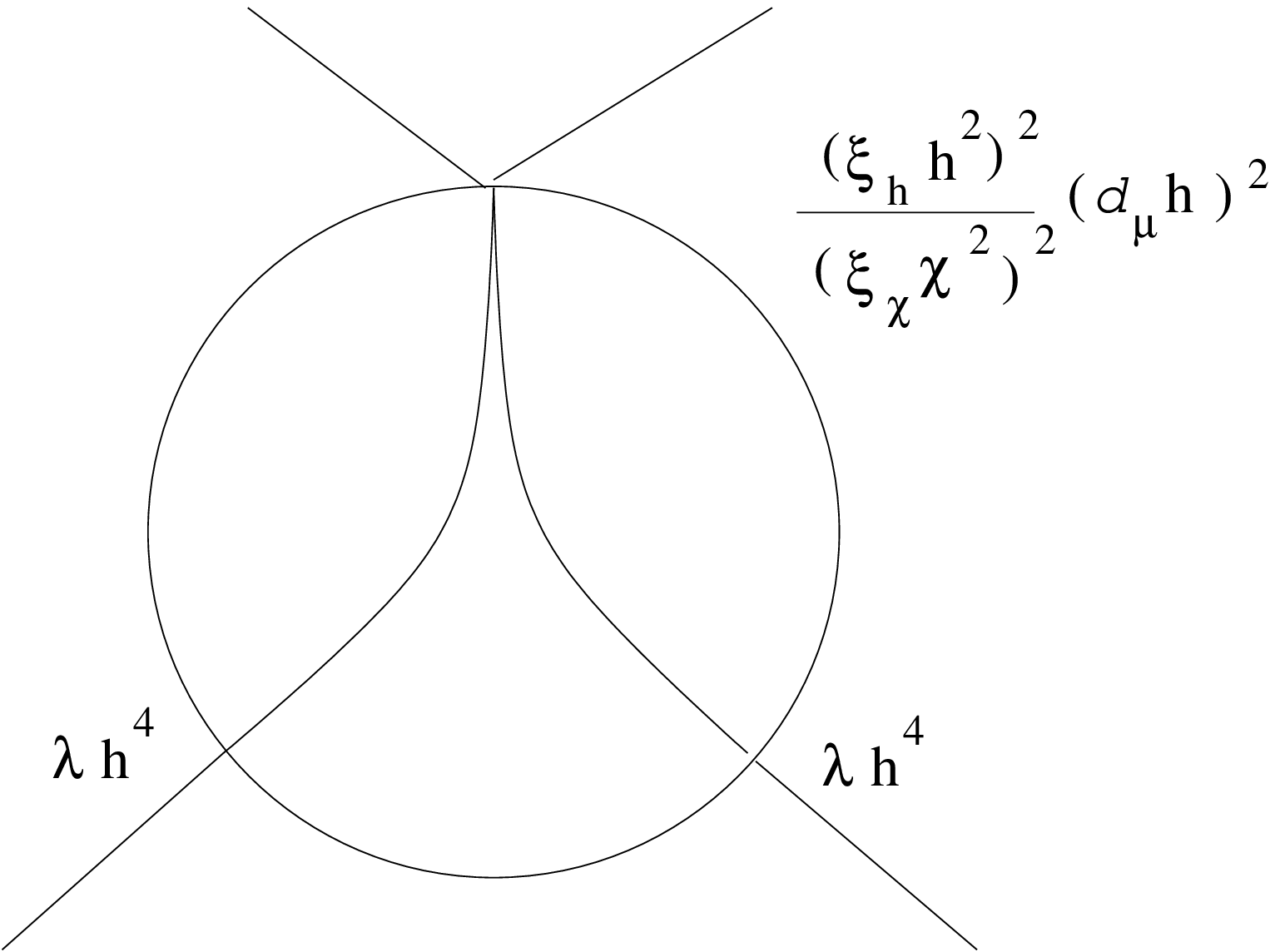}
\caption{A three loop contribution to higher-dimensional operators
that is leading in momenta.}
\label{3loop} 
}
%%%%%%%%%%%%%%%%%%%%%%%%%%%%%%%%%%%%%%%%%%%%%%%%%%%%%%%%
It is therefore necessary to introduce a counterterm with the
structure
\be
 \left(\frac{1}{16\pi^2}\right)^3
\frac{1}{\varepsilon^2}\left(\frac{\xi_h}{\xi_\chi\chi^2}\right)^2\left[\Box
h^2\right]^2~.
\ee
This leads to an estimate of the energy domain of validity of the QSI
effective field theory associated with the action  (\ref{L}):
\be
 E^2 \lesssim \frac{(16\pi^2)^2}{\lambda}
 \frac{\xi_\chi\chi_0^2}{\xi_h},
\ee
which is of the order of the Planck scale. 

The counter-term Lagrangian $L_{ct}$, removing ultraviolet
divergencies, can be made local in the following sense. The
perturbative construction of scale-invariant theories implies the
separation of the quantum fields into a c-number part related to the
background $\chi_0,h_0$, which breaks the QSI, and perturbations,
$\chi=\chi_0 +\delta\chi,~~h=h_0+\delta h$. If $L_{ct}$ satisfies
\be
\left[\frac{\partial}{\partial\chi_0}-\frac{\delta}{\delta(\delta
\chi)}\right]L_{ct}=0~,~~~~\\
\left[\frac{\partial}{\partial h_0}-\frac{\delta}{\delta(\delta h)}\right]L_{ct}=0~,
\label{locality}
\ee
then it will be a local function of the fields $\chi$ and $h$ (i.e.
depend only on the combinations $\chi_0 +\delta\chi$ and $h=h_0+\delta
h$.
The fact that the regularized ($n\neq4$) but non-renormalized Green's
functions of the theory (\ref{L}) with prescription  (\ref{GR})
satisfy the equations (\ref{locality}) ensures that this is indeed the
case.

To conclude, the QSI theories with spontaneous breaking of scale
invariance are predictive at energies $E \ll M_P$ but require an
ultraviolet completion at Planck energies. It is intriguing that in
order to make these theories  compartible with the tests of General
Relativity, the gravity should be introdiced in a scale-invariant way
\cite{Wetterich:1987fk,Wetterich:1987fm,Shaposhnikov:2008xb}. This
leads to derivative coupling of the dilaton to matter fields, evading
the fifth-force bounds on the massless scalars.

{\bf Acknowledgements.} F.T. thanks EPFL for hospitality.  F.T. is
partly supported by the grant of the President of the Russian
Federation NS-7293.2006.2 (government contract 02.445.11.7370). This
work was supported by the Swiss National Science Foundation.

\newpage

\bibliographystyle{/home/mshaposh/D/papers/NotCompleted/HiggsMass/two_loop/final/JHEP.bst}
\bibliography{/home/mshaposh/D/papers/NotCompleted/HiggsMass/two_loop/hep-ph_JHEP/all}

\end{document}